\documentclass[conference]{IEEEtran}
\IEEEoverridecommandlockouts
\usepackage{cite}
\usepackage{amsmath,amssymb,amsfonts}
\usepackage{algorithmic}
\usepackage{graphicx}
\usepackage{textcomp}
\usepackage{xcolor}
\usepackage{physics}
\usepackage{caption}
\usepackage{subcaption}
\usepackage{hyperref}

\usepackage{tikz}
\usetikzlibrary{quantikz2}

\def\BibTeX{{\rm B\kern-.05em{\sc i\kern-.025em b}\kern-.08em
    T\kern-.1667em\lower.7ex\hbox{E}\kern-.125emX}}

\makeatletter 
\newcommand{\linebreakand}{%
  \end{@IEEEauthorhalign}
  \hfill\mbox{}\par
  \mbox{}\hfill\begin{@IEEEauthorhalign}
}
\makeatother 

\begin{document}

\title{CleanQRL: Lightweight Single-file Implementations of Quantum Reinforcement Learning Algorithms}

\author{
\IEEEauthorblockN{Georg Kruse}
\IEEEauthorblockA{\textit{Fraunhofer IISB} \\
\textit{Technical University Munich} \\
Erlangen, Germany \\
georg.kruse@iisb.fraunhofer.de}
\and
\IEEEauthorblockN{Rodrigo Coelho}
\IEEEauthorblockA{\textit{Fraunhofer IISB} \\
Erlangen, Germany \\
rodrigo.coelho@iisb.fraunhofer.de}
\and
\IEEEauthorblockN{Andreas Rosskopf}
\IEEEauthorblockA{\textit{Fraunhofer IISB} \\
Erlangen, Germany \\
andreas.rosskopf@iisb.fraunhofer.de}
\and
\linebreakand
\IEEEauthorblockN{Robert Wille}
\IEEEauthorblockA{\textit{Technical University Munich} \\
Munich, Germany \\
robert.wille@tum.de}
\and
\IEEEauthorblockN{Jeanette Miriam Lorenz}
\IEEEauthorblockA{\textit{Fraunhofer IKS} \\
\textit{Ludwig Maximilian University} \\
Munich, Germany \\
jeanette.miriam.lorenz@iks.fraunhofer.de}
}
\maketitle

\begin{abstract}

At the interception between quantum computing and machine learning, Quantum Reinforcement Learning (QRL) has emerged as a promising research field. Due to its novelty, a standardized and comprehensive collection for QRL algorithms has not yet been established. Researchers rely on numerous software stacks for classical Reinforcement Learning (RL) as well as on various quantum computing frameworks for the implementation of the quantum subroutines of their QRL algorithms. Inspired by the \emph{CleanRL} library for classical RL algorithms, we present \emph{CleanQRL}, a library that offers single-script implementations of many QRL algorithms. Our library provides clear and easy to understand scripts that researchers can quickly adapt to their own needs. Alongside \emph{ray tune} for distributed computing and streamlined hyperparameter tuning, \emph{CleanQRL} uses \emph{weights\&biases} to log important metrics, which facilitates benchmarking against other classical and quantum implementations. The \emph{CleanQRL} library enables researchers to easily transition from theoretical considerations to practical applications.

\end{abstract}


\section{Introduction}

Quantum Reinforcement Learning (QRL) has recently emerged as a promising and popular research field within quantum computing. Its core idea is to leverage the principles of quantum mechanics to improve the performance of classical Reinforcement Learning (RL) agents. In this work, we are particularly concerned with variational QRL, where a Parametrized Quantum Circuit (PQC) replaces the neural network typically used in RL as a function approximator. The result is a hybrid algorithm, where the logic and the optimization of the parameters of the PQC are deferred to a classical computer. In other words, QRL builds upon classical RL algorithms by incorporating quantum subroutines into the original algorithms. Researchers have introduced various quantum versions of many RL algorithms such as REINFORCE \cite{jerbi2021parametrized, sequeira2023policy, meyer2023quantum}, DQN \cite{lockwood2020reinforcement, chen2020variational, skolik2022quantum}, PPO \cite{kwak2021introduction, hsiao2022unentangled, druagan2022quantum} or DDPG \cite{wu2025quantum,schenk2024hybrid, Silvirianti2024}.

The performance of classical RL algorithms depends on numerous implementation decisions: For example, DQN relies on experience replay, which requires a buffer that stores experiences that can be resampled to train the algorithm. It also employs a target network with frozen weights that are updated with a certain frequency to stabilize training, improving performance. Moreover, the exploration-exploitation trade-off can be balanced according to different strategies, and various network architectures offer additional degrees of freedom. These numerous implementation decisions highlight the significant flexibility inherent to RL, which leads to a variety of divergent versions of the very same algorithms. Thus, many RL libraries have been created to guide researchers and users through these decisions and provide them with a robust baseline. Most of these libraries, such as the popular \emph{stable baselines} \cite{raffin2021stable}, opt for a modular approach where each implementation spans across many files and classes. While this allows the user to modify specific parts of the implementation without the need to adapt the remainder of the code, it requires a deep understanding of a large and mostly hidden codebase. In short, these modular approaches can often be self-defeating, since the modularity leads to complex, large and bloated codebases, making even a simple change quite complex in practice.

By contrast, the RL library \emph{CleanRL} \cite{huang2022cleanrl} opts for a different approach. There, each RL algorithm is implemented as a single high-quality script. All required functions and classes are defined within this self-containing script in an easy to understand manner. The user can quickly understand how each algorithm works because all the moving wheels are in plain sight, not hidden away in a giant codebase. This transparency facilitates modifications and customizations compared to modular libraries.

Due to QRL's novelty, no standardized (Q)RL library yet exists. When implementing a QRL algorithm, most researchers choose one of the established RL libraries or rely on custom implementations for the classical part, and then use a quantum machine learning framework, typically either \emph{pennylane} \cite{bergholm2018pennylane}, \emph{qiskit} \cite{javadi2024quantum} or \emph{tensorflow-quantum} \cite{broughton2020tensorflow} to implement the PQC. Consequently, even if researchers publish their code online, no two implementations follow the same foundations. This lack of an established QRL library hinders progress in the field and makes meaningful comparisons difficult.

To address these shortcomings in QRL research, we present \emph{CleanQRL}, a library containing high-quality single-script implementations of many QRL algorithms. Building upon the work of \emph{CleanRL}, we address the issue of fragmented QRL implementations by offering clean, self-contained scripts of the most popular QRL algorithms. For the classical logic, we mainly rely on the implementations of \emph{CleanRL}, while using \emph{pennylane} and \emph{pytorch} for the implementation of the PQCs. However, we also include a classical version of each script such that a quantum-classical comparison is possible.

Our framework incorporates \emph{ray tune} \cite{liaw2018tune} for distributed computing and hyperparameter optimization, alongside  \emph{weight\&biases} \cite{wandb} for visualization and logging. Exemplary configuration files as well as tutorial implementations are provided, allowing users to seamlessly progress from theoretical interest in QRL to practical implementation.

\section{Related Work}

RL libraries can be divided into training libraries and environment libraries. On the one hand, the training libraries contain implementations of many popular RL algorithms, such as DQN and PPO. Some of the most popular ones are \emph{stable baselines} \cite{raffin2021stable}, \emph{ray RLlib} \cite{liang2017ray}, \emph{torchRL} \cite{bou2023torchrl}, \emph{pearl} \cite{zhu2024pearl}, \emph{tianshou} \cite{weng2022tianshou} and \emph{dopamine} \cite{castro2018dopamine}. As previously mentioned, these libraries tend to implement algorithms in a modular manner, where each algorithm spans across many files and classes. Other libraries, such as \emph{CleanRL}  \cite{huang2022cleanrl}, implement each algorithm as a single high-quality script. In order to speed up the training process, other libraries opt for implementations in jax, a high-performance machine learning library, such as \emph{gymnax} \cite{gymnax2022github} and \emph{purejaxRL} \cite{lu2022discovered}.

On the other hand, the environment libraries offer a broad variety of games and (optimization) tasks. The most popular is the successor of the \emph{openai gym} \cite{brockman2016openai} called \emph{gymnasium} \cite{towers2024gymnasium}. Other well established libraries are \emph{pettingzoo} \cite{terry2021pettingzoo}, \emph{dm env} \cite{dm_env2019} and \emph{jumanji} \cite{bonnet2023jumanji}. A critical design requirement for these libraries is the seamless interoperability between training and environment libraries - all environment and training libraries follow the same conventions, such that all environments from all environment libraries work with all algorithms from all training libraries.

One major advantage of these libraries is that they enable consistent evaluations. In particular, the environment libraries mentioned above have been widely used to benchmark classical RL algorithms and more recently QRL algorithms. Besides these environments, other popular benchmark collections, such as the  \emph{open RL benchmark} \cite{Huang_Open_RL_Benchmark_2024}, are available. There, not only various RL algorithms are compared on a diverse set of benchmarking environments, but also \emph{different} implementations of the \emph{same} algorithms are evaluated. However, none of these libraries and benchmark collections contain QRL algorithms nor does a distinct quantum training library for QRL exist. However, as recent works by \cite{bowles2024better} and \cite{lorenz2025systematic} have pointed out, especially quantum computing requires consistent benchmarking efforts for further advancements.
 
\begin{figure*}[ht!]
    \centering
    \includegraphics[width=1.0\linewidth]{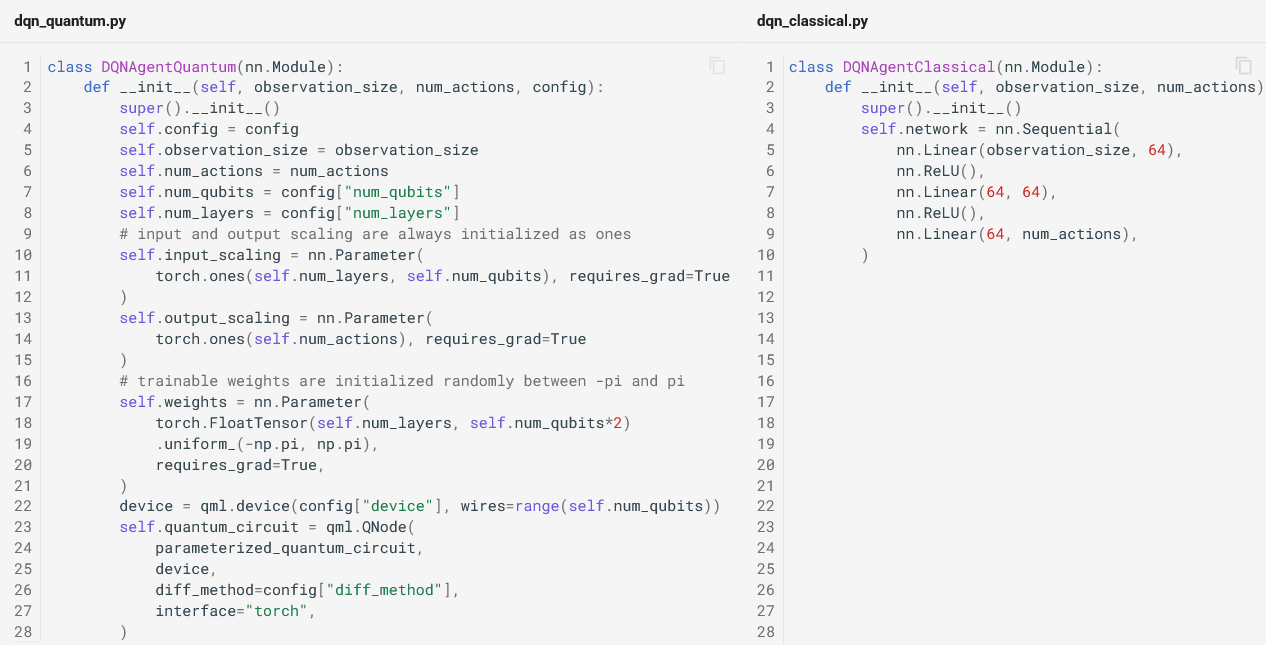}
    \caption{Excerpt of the implementation of the quantum (left) and classical (right) DQN agents. The implementations of the quantum agents require various additional specifications.}
    \label{fig:code_comp}
\end{figure*}

\section{Implementation} \label{implementations}

The proposed QRL library \emph{CleanQRL} is based on the idea that all algorithms are implemented as high-quality single-file scripts. While the main (classical) algorithms' logic follows the implementation decisions of \emph{CleanRL}, there are some key differences. 

First and foremost, \emph{CleanQRL} distinguishes itself by offering both classical and quantum versions of each algorithm. Our implementation aims for as little differences between classical and quantum versions of the algorithms as possible. The key difference is concentrated in the \verb|class Agent|: While the classical agent uses a neural network implemented using the \emph{pytorch} \verb|nn.Module| class, the quantum agent uses a PQC implemented using \emph{pennylane}. The PQC, with its trainable parameters, is defined as a "neural network" using the same \verb|nn.Module| class. This allows for the algorithm's logic and implementation details to remain mainly unchanged, independently of whether the agent is classical or quantum. Both inherit from \verb|nn.Module|, receive the same inputs and return the same outputs; only the function approximator is changed. In Fig. \ref{fig:code_comp}, an excerpt of the comparison between the two implementations can be seen.

We integrate \emph{ray tune} into \emph{CleanQRL} for distributed computing and hyperparameter tuning, providing a simple way to make use of several compute nodes to exhaustively search for the best set of hyperparameters in parallel (see Subsection \ref{sec:tuning}). This is crucial, since the quantum versions of the RL algorithms are highly sensitive to hyperparameters.

\begin{figure*}[h]
\begin{subfigure}{0.19\textwidth}
\includegraphics[width=0.9\linewidth, height=3cm]{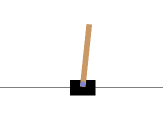} 
\caption{CartPole}
\label{fig:subim1}
\end{subfigure}
\begin{subfigure}{0.19\textwidth}
\includegraphics[width=0.9\linewidth, height=3cm]{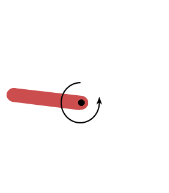}
\caption{Pendulum}
\label{fig:subim2}
\end{subfigure}
\begin{subfigure}{0.19\textwidth}
\includegraphics[width=0.9\linewidth, height=3cm]{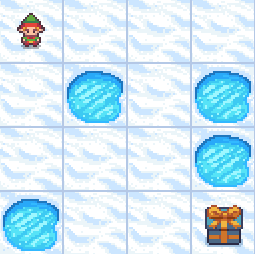}
\caption{FrozenLake}
\label{fig:subim2}
\end{subfigure}
\begin{subfigure}{0.19\textwidth}
\includegraphics[width=0.9\linewidth, height=3cm]{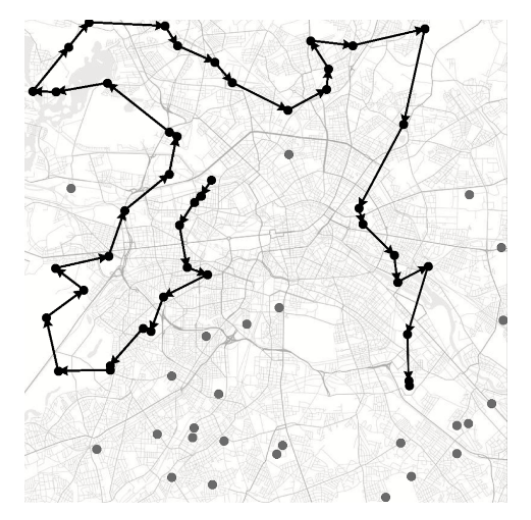}
\caption{TSP}
\label{fig:subim2}
\end{subfigure}
\begin{subfigure}{0.19\textwidth}
\includegraphics[width=0.9\linewidth, height=3cm]{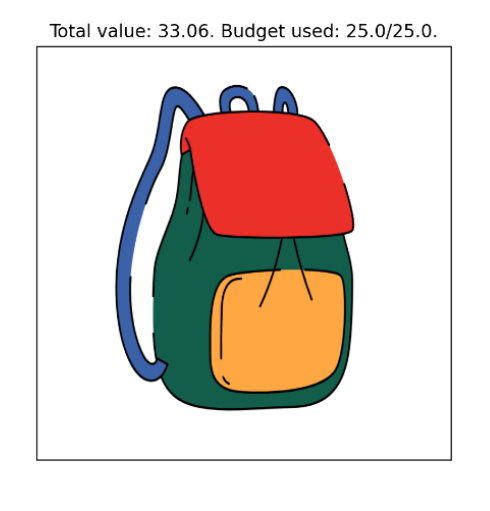}
\caption{Knapsack}
\label{fig:subim2}
\end{subfigure}
\caption{We provide implementations for environments with continuous state space and discrete actions such as CartPole, continuous state space and continuous actions such as Pendulum, and discrete state space and discrete actions such as FrozenLake. Additionally, we provide customized implementations for \emph{jumanji} environments, which use the graph representation of a problem (e.g. TSP) as well as implementations which use the cost Hamiltonian of the problem (e.g. Knapsack).}
\label{fig:envs}
\end{figure*}

Because each script is aimed to be self-contained, there are several scripts for each algorithm depending on the type (classical or quantum) and environment details (see Fig.\ref{fig:envs}). This implementation decision allows for clean scripts at the cost of large amounts of duplicate code. The user needs to select the appropriate script for their respective environment. For example, for the DQN algorithm, the following are some of the available scripts:

\begin{itemize}
    \item \verb|dqn_classical.py|: Follows from the \verb|dqn.py| implementation from \emph{CleanRL}. Works for environments with continuous state and discrete action spaces, which applies to the majority of environments.
    \item \verb|dqn_quantum.py|: The algorithm logic is the same as in \verb|dqn_classical.py| but the neural network is replaced by a PQC.
    \item \verb|dqn_quantum_discrete_state.py|: A quantum version that works for environments with discrete state spaces.
\end{itemize}

Each script follows the same general scheme of three parts:

\begin{itemize}
    \item \verb|def make_env|: The function that initializes the environment.
    \item \verb|class Agent(nn.Module)|: This class inherits from the \emph{pytorch} \verb|nn.Module| class and defines either the classical or the quantum function approximator.
    \item \verb|def training_function|: This function implements the actual (Q)RL algorithm. It loads the important parameters specified in the script or from a configuration file and runs the algorithm.
\end{itemize}

The quantum agents have the following additional classes and functions, which are not required for the classical implementations: 

\begin{itemize}
    \item \verb|class ObservationWrapper|: In RL, the normalization of observations is well established. Due to qubit number limitations and the periodicity of Pauli rotation gates, this is especially important for QRL. We implement an additional \emph{gymnasium} wrapper following the idea of \cite{skolik2022quantum}. The user should modify this wrapper for more sophisticated environments in order to optimize the performance of the quantum agent.
    \item \verb|def parametrized_quantum_circuit|: In this function the PQC ansatz is implemented. We provide  examples of quantum circuits, which are commonly used in recent QRL research.
\end{itemize}

The user can run experiments by executing the appropriate script. At the end of each script, there is a \verb|dataclass| instance specifying the hyperparameters for running the algorithm. With just some small changes to these (or none), the user can immediately run a (Q)RL agent on many \emph{gymnasium}, \emph{jumanji} or custom environments. However, to streamline the development process, we also incorporated running experiments by specifying configuration files, where all hyperparameters are defined (see Subsection \ref{sec:tuning}). For more information on how to set up these configuration files, the reader is referred to the documentation of \emph{CleanQRL}.

Our library works with any \emph{gymnasium} environment and with selected \emph{jumanji} environments. However, in the documentation, we also explain how users can setup custom environments that work with our implementations. Thus, \emph{CleanQRL} either already works or can be easily made to work with any environment that follows the conventions set by \emph{gymnasium}.

Despite our commitment to single file implementations, we recognize that two features, replay buffers and interfaces to \emph{jumanji} environments, present significant challenges to this idea. Combining these elements into single scripts would have resulted in a large, bloated and hard to read code. Hence, we chose to make exceptions to our single-file paradigm for these features.

\subsection{Quantum Circuits}

\begin{figure}[h] 
    \centering
    \includegraphics[width=\linewidth]{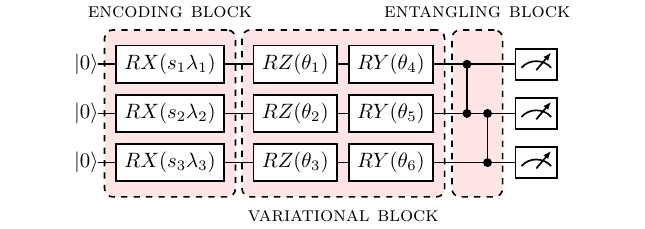}
    \caption{
    A three qubit instance of the ansatz used for a quantum DQN with discrete action space as proposed by \cite{skolik2022quantum}. It has an \emph{encoding block}, where the features of the state $s$ (scaled by trainable parameters $\lambda$) are encoded; a \emph{variational block}, with parameterized quantum gates and trainable weights $\theta$; and an \emph{entangling block}. After $n$ layers of the three blocks, a measurement is performed.} \label{fig:vqc_hea}
\end{figure}

The quantum circuits in each script are based on the ansatzes used in \cite{skolik2022quantum} (see Fig \ref{fig:vqc_hea}) and other recent publications. See the documentation of \emph{CleanQRL} for more details. We use \emph{pennylane} to implement the PQC in combination with the \emph{pytorch} \verb|nn.Module| class. The quantum circuits implemented are called hardware-efficient and serve as an example of an ansatz to use. However, the choice of the ansatz is important for the performance of quantum agents. In particular, hardware-efficient ansatzes are known to suffer from a scalability problem called the Barren-Plateau Phenomenon \cite{larocca2024review}. While the discussion of the cause of this phenomenon is beyond the scope of this work, users should keep in mind that most of the implemented PQC's are just examples of simple and generic ansatzes that may require modifications in order to work with more complex environments. In our framework, such modifications are straightforward to implement. 

\begin{figure*}[h]
\begin{subfigure}{0.49\textwidth}
\includegraphics[width=\linewidth, height=5cm]{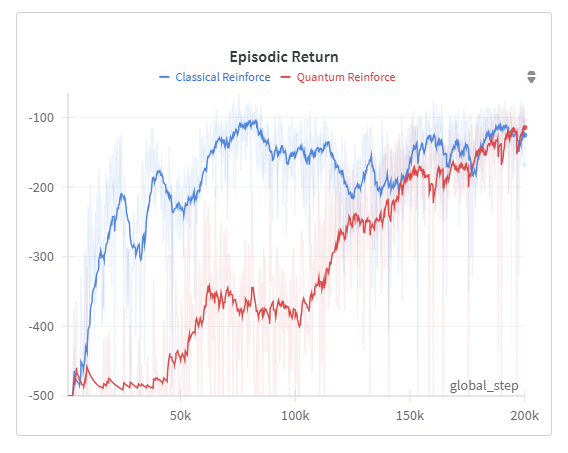} 
\caption{}
\label{fig:reinforce_acrobot}
\end{subfigure}
\begin{subfigure}{0.49\textwidth}
\includegraphics[width=\linewidth, height=5cm]{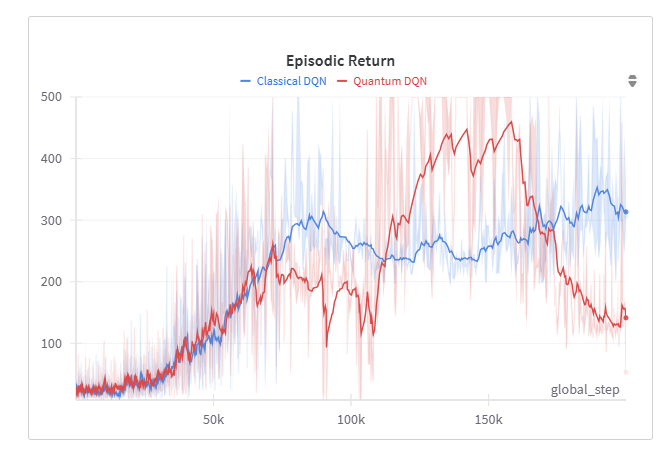}
\caption{}
\label{fig:dqn_cartpole}
\end{subfigure}
\caption{A comparison between the classical and quantum Reinforce/DQN algorithms evaluated on Acrobot-v1/Cartpole-v1 from \emph{gymnasium} (Subfigures \ref{fig:reinforce_acrobot}/\ref{fig:dqn_cartpole}). The screenshot is taken from a \emph{weights\&biases} report. All reports for all implementations can be found in the \emph{CleanQRL} documentation.}
\label{fig:run}
\end{figure*}

\subsection{Hyperparameter Tuning and Logging}\label{sec:tuning}

\emph{CleanQRL} makes use of \emph{ray tune} for distributed computing and hyperparameter tuning, which offers several advantages. On the one hand, quantum models are inherently computationally expensive to simulate, both in terms of processing power and memory. On the other hand, QRL models are usually highly sensitive to hyperparameters and require an extensive search until the desired performance is reached. The use of \emph{ray tune} allows one to make use of several compute nodes, distributing the computational cost of the simulation, and enables multiple training runs in parallel. This speeds up the search for optimal sets of hyperparameters.

Besides the possibility to run the algorithms as single file scripts, we provide four run files according to whether the user wants to use \emph{ray tune} for hyperparameter optimization or not:

\begin{itemize}
    \item \verb|main.py|: This file loads a configuration file and trains a single RL agent without using \emph{ray tune} 
    \item \verb|main_batch.py|: This file loads a list of configuration files and runs each of them sequentially.
    \item \verb|tune.py|: This file loads a configuration file and performs a \emph{ray tune}  hyperparameter optimization based on the specified parameters.
    \item \verb|tune_batch.py|: This file loads a list of configuration files and sequentially performs \emph{ray tune} hyperparameter optimizations for each of them.
\end{itemize}

For logging and data visualization, similarly to \emph{CleanRL}, our framework offers the usage of \emph{weights\&biases}. Several important metrics are logged for each agent and can be easily visualized in \emph{weights\&biases}. Additionally, we also provide custom logging to JSON files as well as \emph{ray tune} logging when performing hyperparameter searches for additional information on CPU and GPU usage, as well as memory utilization.

\section{Tutorials}

The field of QRL is still under active development and novel approaches are published constantly. While we cannot cover all, we provide a set of tutorials for some of the recent publications in QRL. The tutorials also showcase the straightforward integration of novel ideas into the framework and highlight the simplicity of adaptation to custom environments. We provide additional implementations for: 

\begin{itemize}
    \item Graph-based approaches: An interesting idea, which follows the intuition of graph neural networks, is the approach of \cite{skolik2023equivariant}. The authors encode graph-based problems such as the TSP via a customized ansatz which takes the structure of the problem into account. 
    \item Hamiltonian-based approaches: In quantum computing,  algorithms such as QAOA are used to optimize combinatorial optimization problems, which are encoded as Hamiltonians. The work of \cite{kruse2024hamiltonian} shows how to utilize these Hamiltonians as ansatzes for QRL. 
    \item Maze games: Many classes of QRL algorithms like  amplitude amplification-based models or free energy models have been proposed. The few environments on which all these QRL algorithms can be benchmarked are maze games. 
    \item Noise models: Most QRL researchers use state-vector simulators for their experiments and only a very limited number has access to quantum hardware. Noise models can bridge the gap between simulation and hardware and help to understand the potential of novel implementations \cite{skolik2023robustness}.
\end{itemize}

\section{Benchmarks}

For each environment considered (see Fig.\ref{fig:envs} for examples), we provide a \emph{weights\&biases} report with a comparison between the classical and quantum versions of the (Q)RL algorithms applicable to that environment (see Fig.\ref{fig:run} for an example). We also provide a configuration file for each algorithm, which makes it easy to reproduce the results. For all tested environments, we did not perform an extensive hyperparameter search, as this would be extremely expensive in terms of computational resources. We invite the community to contribute by adding even better parameter configuration files to our library for other researchers to explore. In classical RL, the main figure of merit is the sample efficiency in terms of environmental steps to solution. However, for QRL, the high cost of quantum circuit executions needs to be considered. In particular, due to the \emph{parameter-shift rule}, the number of circuit executions needed for gradient estimation depends linearly on the number of parameters of the quantum circuit. Therefore, we create additional reports to illustrate not only sample efficiency, but also efficiency w.r.t. circuit executions of the quantum algorithms. All reports are compatible with the \emph{open RL benchmark}, enabling easy comparison to other classical (and future quantum) libraries.

\section{Conclusion}

In this work, we present \emph{CleanQRL}, a python library that implements many QRL algorithms in single high-quality scripts, following the principles of \emph{CleanRL}. Our framework allows researchers to prototype novel ideas quickly due to clean and easy to understand implementations of many QRL algorithms. Through the integration of \emph{ray tune}, it simplifies distributed computing and speeds up hyperparameter optimization. Via \emph{weights\&biases}, easy logging and visualization is enabled. The latter makes it easy to compare novel implementations with other algorithms provided by the \emph{open RL benchmark}. \emph{CleanQRL} establishes a consistent foundation for QRL algorithms, allowing for more nuanced comparisons either between quantum algorithms or classical and quantum algorithms, ultimately attracting more researchers to contribute to this agile research field.

\section*{{Code Availability and Documentation}}
The code to reproduce the results as well as the data used to generate the plots in this work can be found at https://github.com/fhg-iisb/cleanqrl and tutorials and documentation at https://fhg-iisb.github.io/cleanqrl-docs/.

\section*{{Acknowledgements}}
The research is part of the Munich Quantum Valley, which
is supported by the Bavarian state government with funds from
the Hightech Agenda Bayern Plus.

\bibliographystyle{apalike}
{\small
\bibliography{conference_101719}}

\end{document}